\documentclass[12pt,preprint]{aastex}
\shorttitle{Accurate PSF astrometry and photometry}

\shortauthors{Marois et al.}

\begin{document}

\title{Accurate Astrometry and Photometry of Saturated and Coronagraphic Point Spread Functions}
\author{Christian Marois$^1$, David Lafreni\`{e}re$^2$, Bruce Macintosh$^1$ \& Ren\'{e} Doyon$^2$}
\affil{$^{1}$ Institute of Geophysics and Planetary Physics L-413,\\ Lawrence Livermore National Laboratory, 7000 East Ave, Livermore, CA 94550}
\affil{$^{2}$ D\'{e}partement de physique and Observatoire du Mont M\'{e}gantic, Universit\'{e} de Montr\'{e}al, C.P. 6128, Succ. A,\\ Montr\'{e}al, Qc, Canada H3C 3J7}
\email{cmarois@igpp.ucllnl.org david@astro.umontreal.ca bmac@igpp.ucllnl.org doyon@astro.umontreal.ca} 

\begin{abstract}   
Accurate astrometry and photometry of saturated and coronagraphic point spread functions (PSFs) are fundamental to both ground- and space-based high contrast imaging projects. For ground-based adaptive optics imaging, differential atmospheric refraction and flexure introduce a small drift of the PSF with time, and seeing and sky transmission variations modify the PSF flux distribution. For space-based imaging, vibrations, thermal fluctuations and pointing jitters can modify the PSF core position and flux. These effects need to be corrected to properly combine the images and obtain optimal signal-to-noise ratios, accurate relative astrometry and photometry of detected objects as well as precise detection limits. Usually, one can easily correct for these effects by using the PSF core, but this is impossible when high dynamic range observing techniques are used, like coronagrahy with a non-transmissive occulting mask, or if the stellar PSF core is saturated. We present a new technique that can solve these issues by using off-axis satellite PSFs produced by a periodic amplitude or phase mask conjugated to a pupil plane. It will be shown that these satellite PSFs track precisely the PSF position, its Strehl ratio and its intensity and can thus be used to register and to flux normalize the PSF. A laboratory experiment is also presented to validate the theory. This approach can be easily implemented in existing adaptive optics instruments and should be considered for future extreme adaptive optics coronagraph instruments and in high-contrast imaging space observatories.
\end{abstract}

%\noindent{\em Suggested running page header:} Accurate PSF Astrometry and Photometry

\section{Introduction}
Accurate astrometry and photometry of saturated and coronagraphic point spread functions (PSFs) are fundamental to both ground- and space-based high contrast imaging of brown dwarfs and exoplanets. Achieving precise astrometry is important to minimize the time required to confirm via proper motion analysis that two objects are gravitationally bound, i.e. that a newly found companion is not a background object, and to better constrain its orbit. Precise photometry is essential to better understand the companion physical characteristics. For ground-based adaptive optics imaging, both standard observing techniques and more specialized observing methods, like simultaneous spectral differential imaging \citep{racine1999,marois2000,sparks2002,biller2004,marois2004phd,marois2005}, angular differential imaging \citep{marois2004phd,liu2004,marois2006} and coronagraphy \citep{lyot1932}, generally require the acquisition of a long sequence of images. The object airmass, the telescope orientation and the observing conditions (seeing and sky transmission) change during the sequence, producing point spread function (PSF) flux and Strehl ratio variations over time. The PSF is also slowly moving and evolving due to flexure and differential atmospheric refraction if a near-infrared camera is used while operating the adaptive optics wavefront sensor in the visible. Space-based imaging is also affected by similar problems from vibrations, thermal fluctuations and pointing jitters. If not corrected, these effects bias the relative photometry and astrometry of detected sources and the estimated sensitivity limits. To obtain the optimal signal-to-noise ratio (S/N) and precise astrometry and photometry after combination of all acquired images, each individual image needs to be accurately registered and flux normalized.

Standard techniques to properly register a PSF and to estimate the relative astrometry and photometry generally require the PSF core. If the PSF core is saturated to improve observing efficiency or if a non-transmissive focal plane occulter is used, the PSF core is not visible and other techniques need to be applied. If available, an off-axis sharp ghost image can be used to register the PSF and estimate its core position, but the possibility of a non-common motion between the ghost and the PSF makes the registering uncertain. 

The usual technique for estimating sensitivity limits (or a source relative intensity) when the PSF core is saturated or occulted is to compare the noise in the combined image (or a source peak intensity) to the peak intensity of an unsaturated/unocculted PSF acquired before and/or after the saturated/occulted image sequence. Such estimates can be biased by the short exposure time and the generally small number of unsaturated/unocculted images. Furthermore, the observing conditions of the unsaturated/unocculted images can be significantly different from those of the saturated/occulted image sequence.

To obtain the highest S/N ratio in the combined image, one would want, ideally, to normalize the images such that any point source, after image registration, has the same number of counts per full width at half maximum (FWHM) in each image. The images would then be weighted by the inverse square of their noise and combined.

A simple solution to these astrometric and photometric problems exists. A periodic phase or amplitude mask can be introduced at a pupil plane to produce fainter off-axis copies of the primary PSF. The position and relative intensity of these satellite PSFs are mainly fixed by the mask amplitude and spatial periodicity. These satellite PSFs track precisely the PSF core position as well as its intensity variations. They can thus be used for accurate astrometry and photometry. First, this technique will be analyzed using a simple analytical model and numerical simulations. The technique will then be validated using a laboratory experiment. A technique similar to the one presented in this paper has been independently developed by \citet{siv2006}.

\section{Non-Coronagraphic PSFs\label{pupaberr}}
The effect of a pupil-plane conjugated aberration can be estimated using a far-field approximation. The intensity $I(\eta,\xi)$, where $\eta$ and $\xi$ are coordinates in the focal plane, of the PSF is simply the Fourier transform FT of the complex electric field \citep{schroeder1987} at the pupil

\begin{equation}
I(\eta,\xi) = \left| \rm{FT}(A(x,y) e^{i[\phi(x,y)]}) \right|^2 \equiv \left| F_a \right|^2 \equiv I_a\label{eq1}
\end{equation}

\noindent where $A$ is the wavefront amplitude, equal to one inside the pupil and zero elsewhere, and $\phi$ its phase, having a noise RMS $\sigma$ inside the pupil. Both $A$ and $\phi$ are expressed in coordinates $x$ and $y$ in the pupil and are real functions. The phase $\phi$ has zero mean. The functions $I_a$ and $F_a$ are respectively the aberrated PSF intensity and its complex focal plane electric field. The next two sub-sections will study the effect of adding a specific phase or amplitude pupil-plane mask to the aberrated wavefront. It is assumed in the following analysis that $\phi(x,y)$ is small (space-based or AO imaging), producing diffraction limited images.

\subsection{Pupil-Plane Periodic Phase Mask\label{phasemask}}
If a phase aberration $\phi_{\rm{mask}}$ having a FT equal to $\Phi_{\rm{mask}}$ and a standard deviation $\sigma_{\rm{mask}}$ is introduced on the wavefront at a pupil plane, the modified PSF intensity can be deduced from Eq.~\ref{eq1}. The aberration $\phi$ is simply replaced by $\phi + \phi_{\rm{mask}}$ and the following expression is found

\begin{equation}
I(\eta,\xi) = \left| \rm{FT}(A(x,y) e^{i[\phi(x,y)+\phi_{\rm{mask}}(x,y)]}) \right|^2.\label{eq2}
\end{equation}

\noindent Separating the effect of the mask, we find

\begin{equation}
I(\eta,\xi) = \left| \rm{FT}([A(x,y) e^{i[\phi(x,y)]}][e^{i[\phi_{\rm{mask}}(x,y)]}]) \right|^2.\label{eq3}
\end{equation}

\noindent Following the work of \citet{bloemhof2001,siv2002,perrin2003}, assuming $\phi_{\rm{mask}}$ is small, the complex exponential of $\phi_{\rm{mask}}$ can be expanded using a Taylor approximation to find

\begin{equation}
I(\eta,\xi) \cong \left| \rm{FT}([A(x,y) e^{i[\phi(x,y)]}][1+i\phi_{\rm{mask}} -\frac{\phi^2_{\rm{mask}}}{2}+\ldots]) \right|^2.\label{eq4}
\end{equation}

\noindent We can then use the equation derived in \citet{perrin2003} to find 

\begin{equation}
I(\eta,\xi) \cong \sum_{n=0}^{\infty} p_n
\end{equation}

\begin{equation}
p_n = i^n \sum_{k=0}^{n}\frac{(-1)^{n-k}}{k!(n-k)!}(F_a \star^k \Phi_{\rm{mask}})(F_a^* \star^{n-k} \Phi_{\rm{mask}}^*).\label{perrin}
\end{equation}

\noindent If the phase mask is chosen to be a periodic function, $\Phi_{\rm{mask}}$ is sharply peaked at symmetric locations. The terms proportional to $F_a \star^i \Phi_{\rm{mask}}$ thus simply add replicas of $F_a$ at each peak of $\Phi_{\rm{mask}}$ and at integer multiples of these peaks for multiple convolutions. Keeping only terms up to the second order, we find

\begin{equation}
I(\eta,\xi) \cong I_a + 2\Im [F^*_a (F_a \star \Phi_{\rm{mask}})] - \Re [F_a^* (F_a \star \Phi_{\rm{mask}} \star \Phi_{\rm{mask}})]+|F_a \star \Phi_{\rm{mask}}|^2\label{eq5}
\end{equation}

\noindent where the symbols $\Im$ and $\Re$ denote respectively the imaginary and real parts. Both the second and third terms are interference effects between the aberrated PSF and $\Phi_{\rm{mask}}$. For the replicas to be greater than $I_a$ at the same separation, $\sigma_{\rm{mask}}$ needs to be larger than the noise component of $\sigma$ at the spatial frequency corresponding to the mask periodicity. Nothing that $F_a$ sharply drops with angular separation, since both the second and third terms are multiplied by $F_a$ and since $\Phi_{\rm{mask}}$ is selected to be brighter than $F_a$ ($\Phi_{\rm{mask}} \gg F_a$ at $\Phi_{\rm{mask}}$ maxima), both terms can be neglected when compared to the fourth one. In that case, we find that the PSF intensity evolution due to the phase mask is simply

\begin{equation}
I(\eta,\xi) - I_a \cong |F_a \star \Phi_{\rm{mask}}|^2.\label{eq6}
\end{equation}

\noindent Since $F_a$ is the aberrated PSF electric field, the periodic phase mask produces replicas of the aberrated PSF, we call these satellite PSFs. Since the satellite PSFs are produced by a convolution of the aberrated PSF, they track the PSF core position and intensity variations. The intensity and location of the satellite PSFs are respectively set only by the amplitude and periodicity of the phase mask, they are thus insensitive, to first order, to a mask offset. A discussion on the neglected terms and their effects on the astrometric and photometric precision can be found in sect.~\ref{prec0}.

\begin{figure}
\epsscale{1}
\plotone{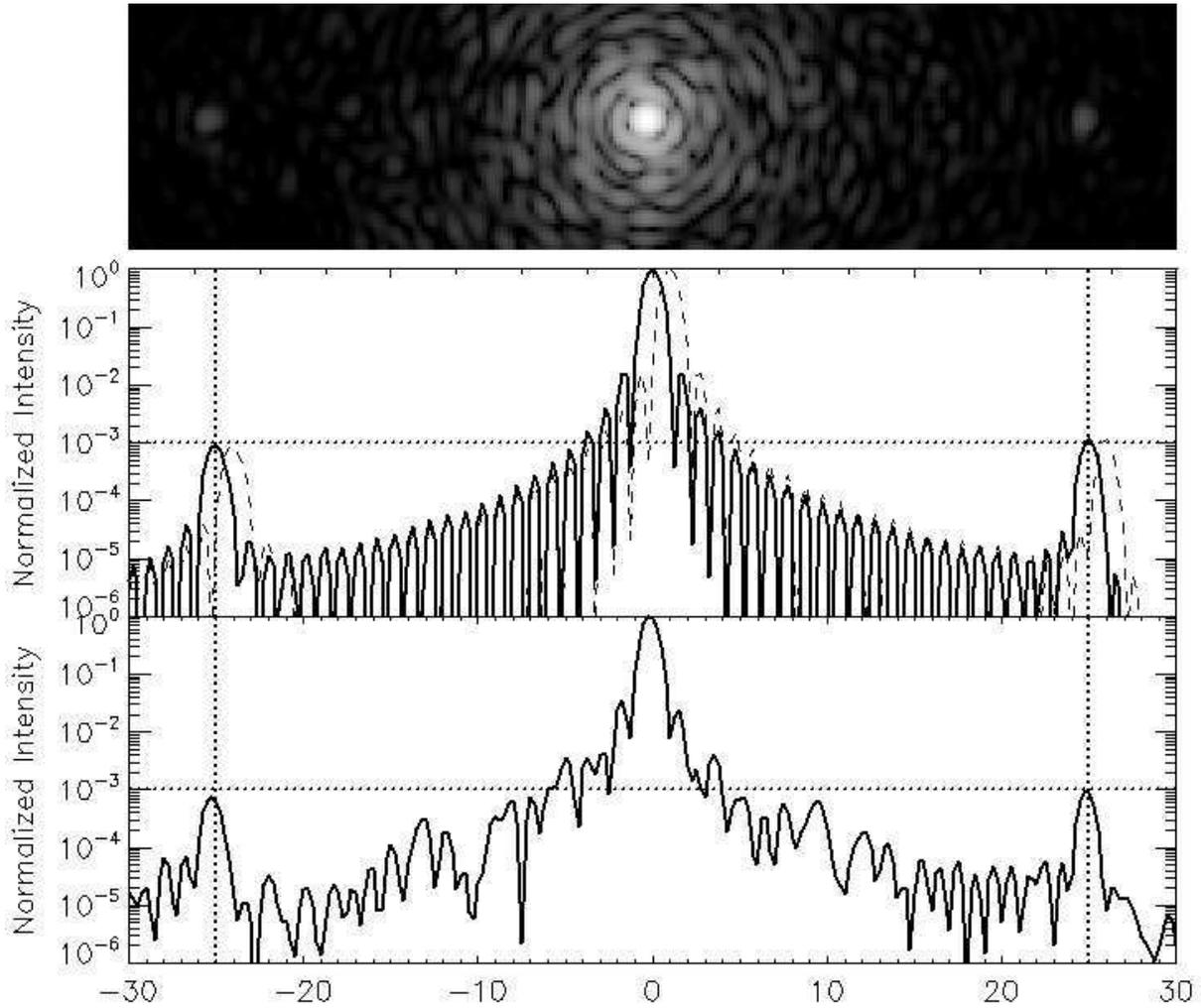}
\caption{Central plot: Simulated monochromatic PSF with added sine wave phase mask (solid line). The dashed line shows the same simulation with an added tilt to show that the satellite PSFs track the PSF core position. Top panel and bottom plot: the same simulation with an added 150~nm RMS phase aberration. Dotted lines show the satellite PSFs expected positions and intensities. Top panel is shown on a logarithmic scale.\label{fig1}}
\end{figure}

If we assume that the phase mask is a sine wave, two satellite PSFs will be produced at an angular separation equal to the number of cycles per pupil in $\lambda/D$ units and with a relative intensity equal to $\sigma_{\rm{mask}}^2/2$. Fig.~\ref{fig1} shows an unaberrated 1.6~$\mu $m monochromatic PSF with a 11~nm amplitude sine wave phase mask having 25 cycles per pupil as well as a PSF having the same sine wave phase mask but with a 150~nm RMS phase aberration. The aberration power spectral density (PSD) follows a power-law of index -2.6. It is expected that satellite PSFs will be 1000 times fainter than the PSF and located at 25 $\lambda/D$. PSFs are simulated with a 256 pixel diameter circular pupil inside a $1024\times 1024$ pixel image producing a PSF having 4~pixels per $\lambda/D$. The satellite PSFs are clearly at their predicted location, they have the estimated relative intensity and they track the PSF core position, Strehl ratio and its intensity. These satellite PSFs can thus be used for registration and photometric calibration (see sect.~\ref{prec}).

\subsection{Pupil-Plane Periodic Amplitude Mask}
A similar analysis can be done for a pupil plane conjugated amplitude mask. Introducing an amplitude mask $\epsilon_{\rm{mask}}$ having a FT equal to $E_{\rm{mask}}$ and a standard deviation $\sigma_{\rm{mask}}$ on the wavefront conjugated to the pupil plane modifies Eq.~\ref{eq1} as follows

\begin{equation}
I(\eta,\xi) = |\rm{TF}([A+A\epsilon] e^{i[\phi(x,y)]})|^2.\label{eq7}
\end{equation}

\noindent Reorganizing this equation, we find

\begin{eqnarray}
I(\eta,\xi) = [\rm{TF}(A e^{i[\phi(x,y)]})+\rm{TF}([A\epsilon]e^{i[\phi(x,y)]})][\rm{TF}(A e^{i[\phi(x,y)]})+\rm{TF}([A\epsilon]e^{i[\phi(x,y)]})]^*.\label{eq8}
\end{eqnarray}

\noindent Knowing that the FT of a multiplication is the convolution of the two FTs, we find

\begin{equation}
I(\eta,\xi) = I_a + 2 \Re[F_a(F_a^*\star E^*)]+|F_a\star E|^2.\label{eq9}
\end{equation}

\noindent Following the argument of section~\ref{phasemask}, the second term can be neglected and we find the modified PSF intensity to be

\begin{equation}
I(\eta,\xi) - I_a \cong |F_a\star E|^2.\label{eq10}
\end{equation}

\noindent The satellite PSFs are thus again copies of the primary aberrated PSF. See sect.~\ref{prec0} for a discussion about the neglected term and its effect on the astrometric and photometric precision.

\begin{figure}
\epsscale{1}
\plotone{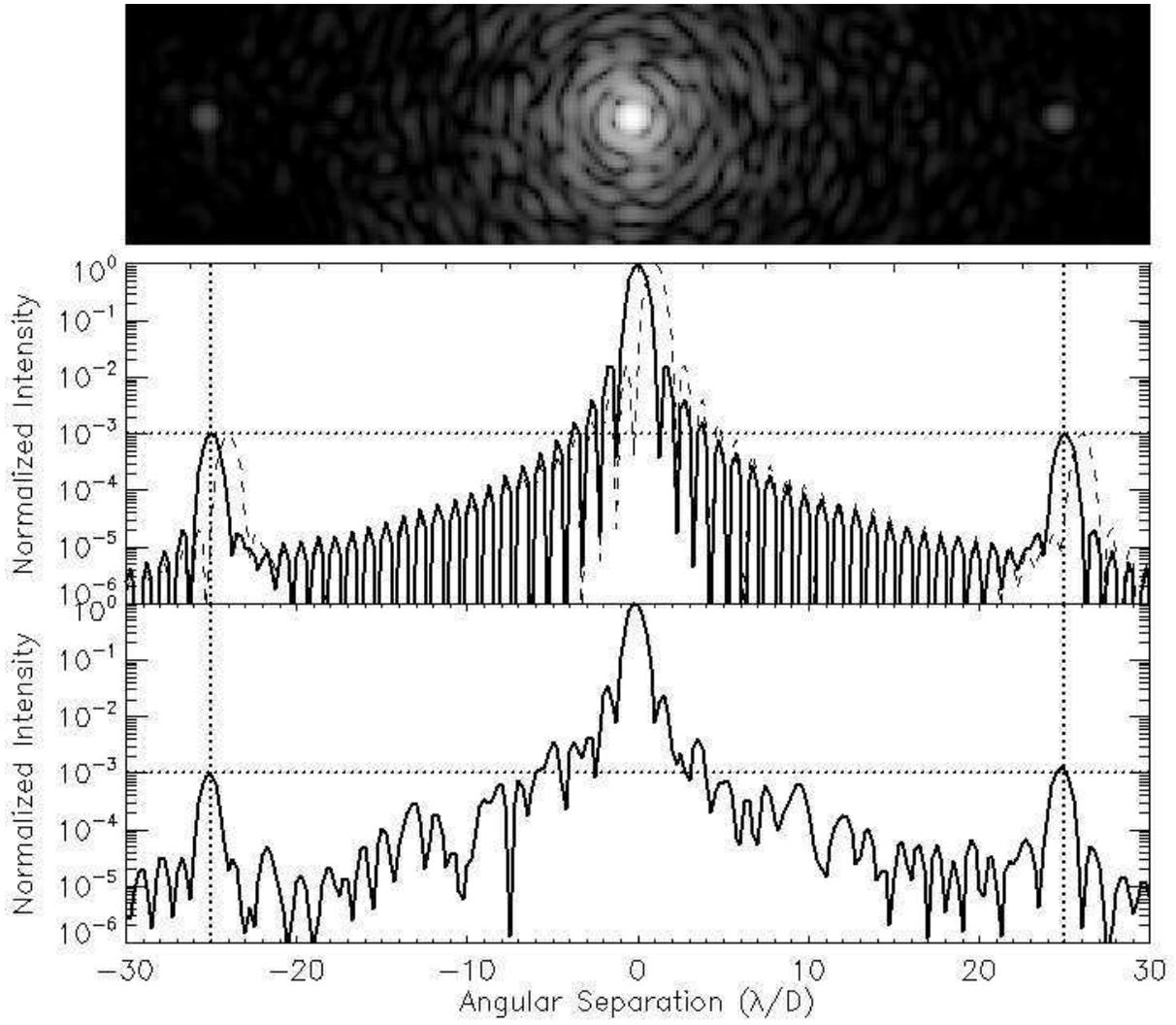}
\caption{Same as Fig.~\ref{fig1} for an added amplitude sine wave mask.\label{fig2}}
\end{figure}

If the amplitude mask is a sine wave, the satellite PSFs have a relative intensity equal to $\sigma_{\rm{mask}}^2/2$ and their separation is equal to the number of cycles per pupil in $\lambda/D$ units. Fig.~\ref{fig2} shows an unaberrated 1.6~$\mu $m monochromatic PSF with a 4\% amplitude sine wave transmission mask having 25 cycles per pupil as well as an aberrated PSF with the same sine wave mask but with 150~nm RMS phase aberration as in section~\ref{phasemask}. It is expected that satellite PSFs will be at 25 $\lambda/D$ separation and 1000 times fainter than the PSF. Similarly to what has been found for a phase mask, the satellite PSFs are at their predicted location, they have the estimated relative intensity and they track the PSF core position and its intensity.

Instead of a sine wave amplitude mask, another option is to simply use a transmission amplitude grating consisting of wires to mask a periodic section of the pupil. In that case, Eq.~\ref{eq10} is still valid and the satellite PSFs will be the convolution of the wire grid Fourier transform with the aberrated PSF. The wire grid can be treated as a multi-slit Young experiment and the satellite PSFs intensity function $I^s$ is thus simply \citep{hecht}

\begin{equation}
I^s(\theta) = I_0^s \left( \frac{\sin \beta}{\beta}\right)^2 \left( \frac{\sin N\alpha}{\sin \alpha} \right)^2 \label{young}
\end{equation}

\noindent where $\theta$ is the angular separation along the satellite PSFs axis, $I_0^s$ is the intensity at $\theta = 0$, $N$ is the number of wires across the pupil and $\beta$ and $\alpha$ are respectively equal to $(kb/2) \sin \theta$ and $(ka/2) \sin \theta$. The constant $k$ is the propagation number ($2 \pi/\lambda$) and the values $a$ and $b$ are respectively the wire spacing and thickness. The first parenthesis in Eq.~\ref{young} is the intensity envelope produced by the diffraction of a single wire (sinc function having $\sim \lambda/b$ FWHM) while the second parenthesis is the intensity modulation from the interference of diffracted light from all wires. Satellite PSFs (local maxima) appear at $m \lambda/a$ separations, where $m$ is an integer. In contrast to the phase and amplitude sine wave masks that show two satellite PSFs, the transmission amplitude grating produces numerous satellite PSFs. The relative intensity, at $\theta = 0$, of the satellite PSFs is equal to the square of the ratio of the area masked by the grating over that in the unobscured pupil region

\begin{equation}
\frac{I^s_0}{I_0} = \left( \frac{\int_{\rm{Pupil}}(A \epsilon )dxdy}{\int_{\rm{Pupil}}(A+A \epsilon)dxdy} \right)^2
\end{equation}

\noindent where $I_0$ is the aberrated PSF peak intensity. For a uniform pupil having numerous wires, the circular shape of the pupil can be neglected and the intensity ratio is approximated by

\begin{equation}
\frac{I^s_0}{I_0} \cong \left( \frac{b}{a-b} \right)^2.
\end{equation}

\noindent If $a \gg b$, the ratio is simply $(b/a)^2$.

\begin{figure}
\epsscale{1}
\plotone{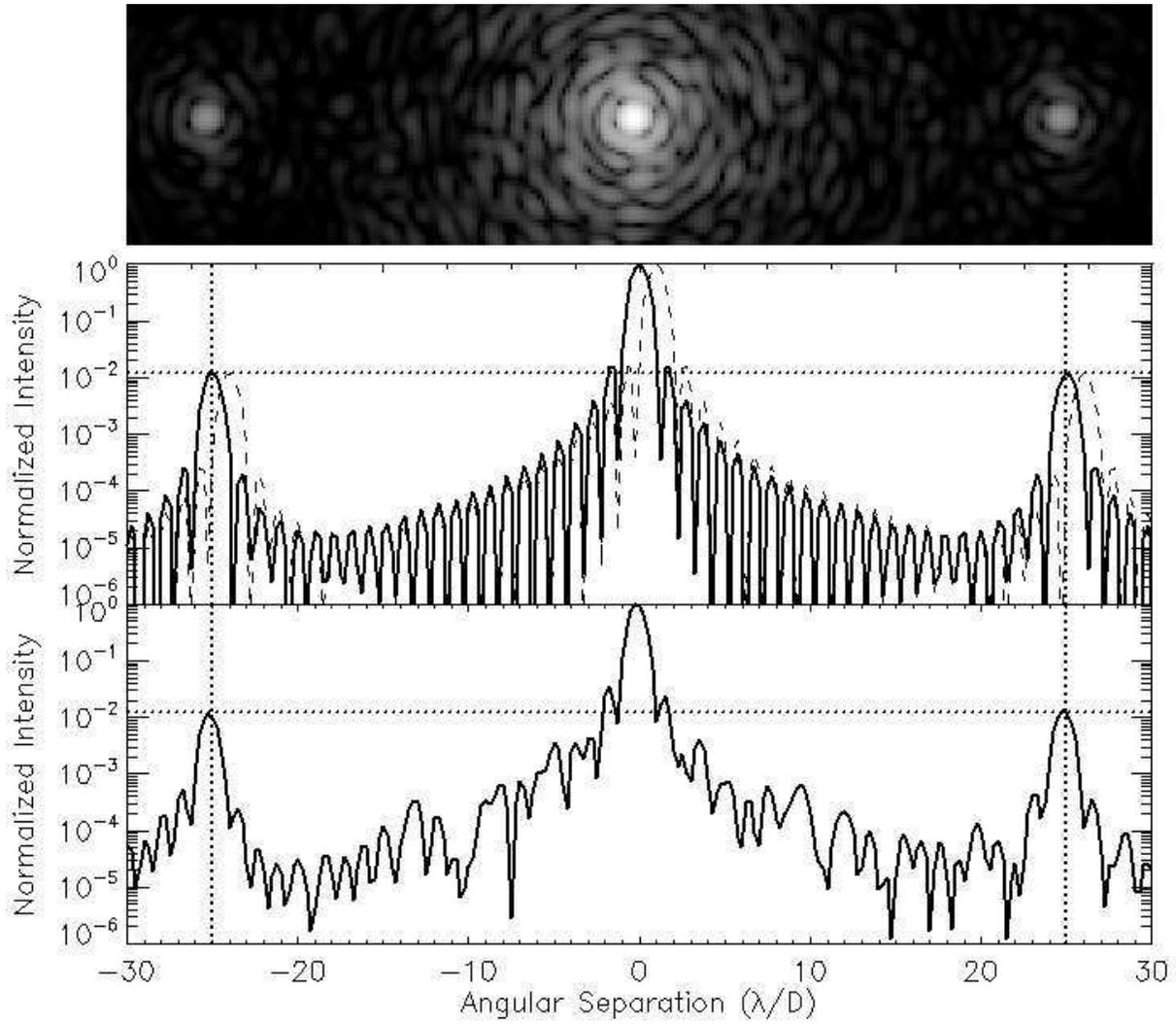}
\caption{Same as Fig.~\ref{fig1} for an added transmissive amplitude grating. This figure shows the first two bright satellite PSFs. Numerous fainter satellite PSFs exist at wider separation.\label{fig3}}
\end{figure}

Fig.~\ref{fig3} shows an unaberrated 1.6~$\mu $m monochromatic PSF simulation with a grating having 25 wires across the pupil, each of width equal to 1/250 of the pupil diameter, as well as an aberrated PSF (150~nm RMS phase aberration) with the same grating (for these simulations, the pupil diameter is 250 pixels and the image has 1000$\times 1000$ pixels). It is expected that the first satellite PSFs will be at 25 $\lambda/D$ separation and 81 times fainter than the PSF. The added satellite PSFs are at their predicted location, they have the estimated relative intensity and they again track the PSF position, Strehl ratio and its intensity (see Sect.~\ref{prec}).

\section{Coronagraphic PSFs with added phase/amplitude mask}
\begin{figure}
\epsscale{1}
\plotone{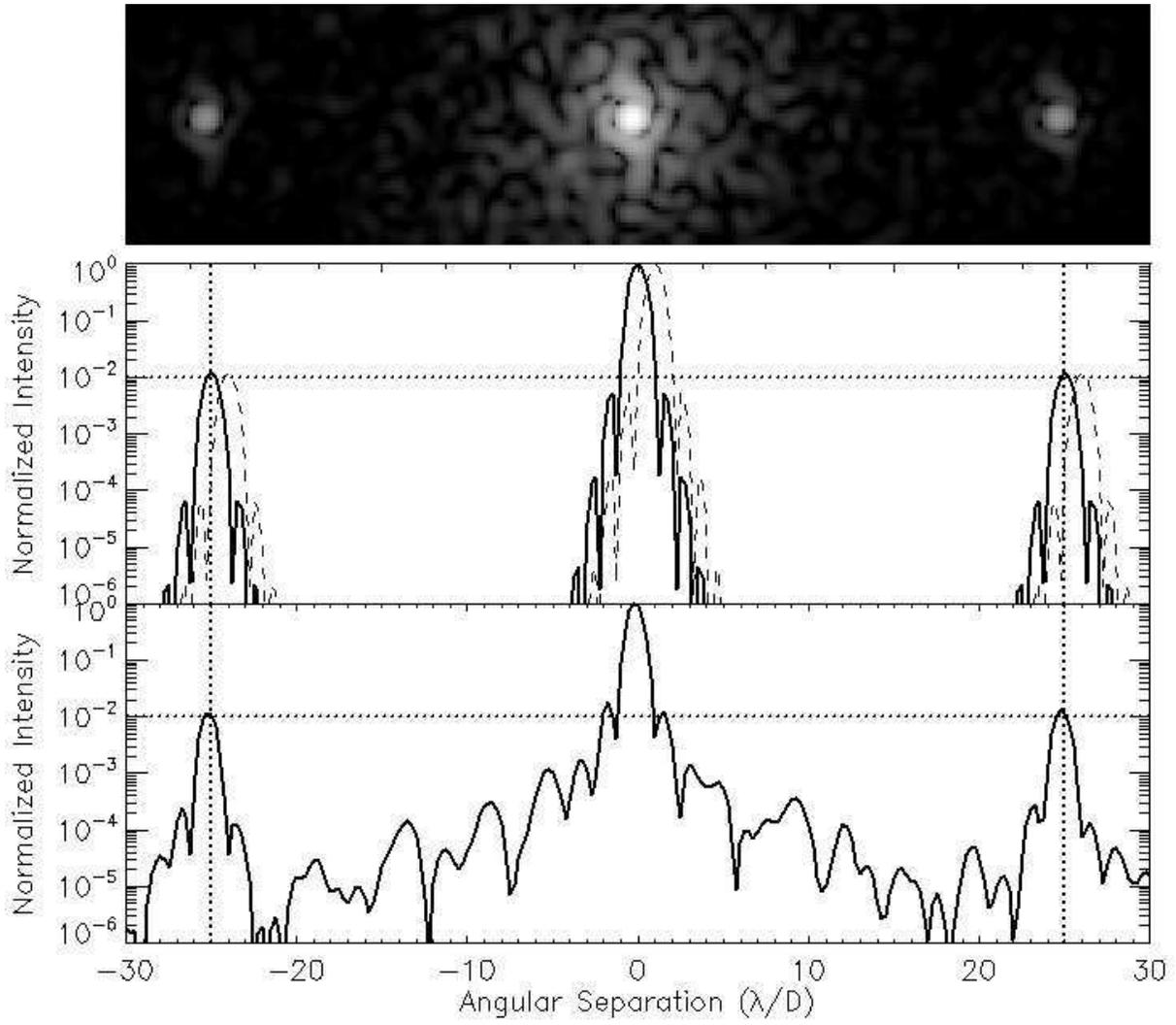}
\caption{Same as Fig.~\ref{fig1} for a Gaussian apodized pupil coronagraph with a transmissive amplitude grating.\label{fig4}}
\end{figure}

The analysis presented in Sect.~\ref{pupaberr} was done using a far field approximation without the use of a coronagraph. If a coronagraph is used, the amplitude/phase mask needs to be introduced before the focal plane occulter so that the satellite PSFs be replicas of the unocculted PSF. If we consider a simple pupil plane Gaussian apodizer, it is easy to understand from Eq.~\ref{eq6} and Eq.~\ref{eq10} that the conclusions of Sect.~\ref{pupaberr} are still valid for a coronagraph since the PSFs are simply obtained with $A$ being the apodized pupil instead of a uniform pupil. The apodized Gaussian coronagraph was chosen here since the PSF diffraction core is still detectable and can thus be used to verify that satellite PSFs track the PSF core position and its intensity. The result would be the same for a band-limited \citep{kuchner2002} or a standard Lyot coronagraph since the grating is introduced in a pupil plane before the focal plane mask occulter and because, similarly to the apodized pupil coronagraph, these coronagraphs produce PSFs which are dominated by the second order halo term \citep{perrin2003,bloemhof2004}. A Gaussian apodized pupil for a 1.6~$\mu $m monochromatic PSF was simulated using the transmissive amplitude grating with and without 150~nm RMS phase aberration (see Fig.~\ref{fig4}). The pupil apodization was simulated by convolving the uniform pupil by a Gaussian having a FWHM equal to one quarter of the pupil diameter. The simulation confirms that the satellite PSFs do again track the PSF core position and intensity.

\section{Expected Astrometric and Photometric precision \label{prec0}}
The aberrated PSF, the neglected terms of Eq.~\ref{eq5} and \ref{eq9} and the rest of the terms in Eq.~\ref{perrin} show symmetries or antisymmetries that can, in the presence of aberrations or mask shifts, slightly affect the PSF center and intensity derived from the satellite PSFs. The task of finding the dominant terms in different regimes is complicated by the fact that $F_a$ is complex; the imaginary and real parts of terms in Eq.~\ref{eq5} and \ref{eq9} are not necessarily respectively antisymmetric and symmetric as in the unaberrated case.

For simplicity let us first consider the case without aberrations. In this case, $F_a$ is the field strength of a perfect PSF; it is thus real and symmetric. The first neglected term of Eq.~\ref{eq5} is antisymmetric and affects the astrometric precision and the second one is symmetric and affects the photometric precision (higher order neglected terms from the Taylor expansion have a negligible effects since they are at least $\Phi_{\rm{mask}}$ times smaller than these two terms). The amplitudes of these terms are function of $\Phi_{\rm{mask}}$. If $\Phi_{\rm{mask}}$ is antisymmetric, the second term is zero and the first term is maximal (astrometric error only), while if $\Phi_{\rm{mask}}$ is symmetric, the first term is zero and the second term is maximal (photometric error only). The symmetry of the mask is determined by its position in the pupil and may vary due to flexure or removal/reinsertion. The astrometric error is proportional to the ratio of the amplitude of the first neglected term to the fourth term of Eq.~\ref{eq5}, or $F_a/\Phi_{\rm{mask}}$. The photometric error is proportional to the ratio of the amplitude of the second neglected term to the fourth term, or $F_a$. Thus astrometric measurements with brighter satellite PSFs are less affected by a mask offset.

For the amplitude mask without aberration, the neglected term of Eq.~\ref{eq9} produces only a symmetric structure (photometric error only). If the mask is antisymmetric, the term is zero, the satellite PSF intensity is thus equal to the expected value, while if the mask is symmetric, the term is maximal and a photometric error is present. Again, a mask offset can induce symmetry variations and cause a photometric error. The photometric precision is proportional to the ratio of the neglected term to the third term of Eq.~\ref{eq9}, or $F_a/\Phi_{\rm{mask}}$. The photometric precision is less sensitive to a mask offset when the satellite PSFs are brighter.

As mentioned previously, since $F_a$ is complex, the case with aberrations is much more complicated. Astrometric and photometric errors can arise from both imaginary and real parts. Furthermore, the residual symmetric and antisymmetric speckle noise from the PSF will also influence the level of accuracy. Intuitively, it is expected that brighter satellite PSFs will result in a better astrometric and photometric precision.

\section{Simulated Performances\label{prec}}
In previous sections, it was shown that periodic amplitude or phase masks can be used for astrometry and photometry calibration of saturated and coronagraphic PSFs. In realistic conditions, the satellite PSFs are overlaid over a background of speckle noise that can affect the determination of the satellite PSFs centers and intensities. It is thus important to adjust the intensity of the satellites PSFs to ensure that the desired accuracy is realized. Since we have assumed that the speckle noise is limiting the astrometric and photometric precision, the sky background, detector read and photon noises are neglected. Numerical simulations are now used to analyze the registering and photometric precision of the technique. Each satellite PSF is cross-correlated with a Gaussian to determine its precise location. The average X and Y positions of two symmetric satellite PSFs are then used to derive the PSF center. This derived position is then compared to the calculated PSF center using the same cross-correlation algorithm on the PSF core. Relative intensities are simply obtained by taking the ratio of the satellite's peak intensity over that of the primary PSF. The coronagraph was simulated by simply convolving the pupil by a Gaussian having a FWHM equal to one quarter of the pupil diameter. The brightness of the satellite PSFs was adjusted to 0.1\% of the PSF peak intensity, which correspond to 100 times the background speckle noise of an aberrated (150 nm RMS) non-coronagraphic PSF. Both the phase and amplitude sine phase masks are antisymmetric on the pupil while the grating is symmetric (it is impossible to get a purely antisymmetric mask using a transmission amplitude grating). The first simulation is performed without aberration (see Table~\ref{tab0}). As mentioned in sect.~\ref{prec0}, only the phase mask shows astrometric errors when the mask is displaced while all mask types show small photometric errors. The use of a coronagraph attenuates interference terms and yields a better astrometric and/or photometric precision in all cases.

Twenty simulations using independent 150~nm RMS phase aberrations were then realized for phase sine, amplitude sine and amplitude grating masks for both cases with and without the coronagraph. Table~\ref{tab1} summarizes the achievable registering and photometric precision. Typically, an astrometric error of 0.3 pixels RMS ($\sim $1/15 $\lambda/D$) is found for both the non-coronagraph and the coronagraph cases. Typical relative intensities are precise to less than 5\%, enough for accurate flux normalization.

Table~\ref{tab2} presents the registering and photometric accuracies for the non-coronagraph and coronagraph cases using the sine wave amplitude mask for different satellite PSF intensities relative to the local speckle noise. Satellite PSFs $\sim $10 times brighter than the speckle noise limit the PSF registering accuracy to 1.2 pixels RMS (1/3 $\lambda/D$) and the photometric calibration to 15-20\%, while satellite PSFs being $\sim $100 times brigther than the background speckle noise limit the PSF registering to 0.3 pixel RMS ($\sim $1/15 $\lambda/D$) and the photometric calibration to 5\%. We note that both the astrometric and photometric precision increase as the square root of the ratio of the satellite PSF intensity to the speckle noise background, i.e. $F_a/\Phi_{\rm{mask}}$ if both the PSF and mask are dominated by the second order halo term at the satellite PSF separation. Since this is the dependance that was found in sect.~\ref{prec0} for the unaberrated case, it is tempting to conclude that the second term of Eq.~\ref{eq9} introduces both astrometric and photometric errors, as was expected from the fact the $F_a$ is complex.

A satellite PSF relative intensity of $\sim 100$ times the background speckle noise seems to be a good choice to avoid over-luminous satellite PSFs while offering good registering and photometric accuracy (0.3 pixel RMS ($\sim $1/15 $\lambda/D$) and 5\%). Such an astrometric precision (0.003 arcsec for diffraction limited H-band images acquired with a 10-m telescope) is sufficient to confirm proper motion of nearby systems (typically 0.01 to 0.1 arcsec/year) in one year.\footnote{For speckle noise limited companion detections, the derived astrometric precision from Sect.~\ref{prec} simulations also mean that companions need to be at least 100 times brighter than the background speckle noise to obtain 1/15~$\lambda/D$ center determination accuracy and allow their proper motion follow-up in one year for typical nearby systems - fainter companions will require a bigger time interval or longer integration time, assuming that the object S/N is increasing with integration time.} Due to detector dynamic range limits, to minimize satellite PSF intensities, it might be more convenient to use a large number of fainter satellite PSFs to average the speckle noise bias, albeit at the cost of a greater FOV contamination, or choose a mask frequency to move the satellite PSFs to wide separations, where the speckle noise is generally smaller.

\section{Laboratory Experiment}
\begin{figure}
\epsscale{1}
\plotone{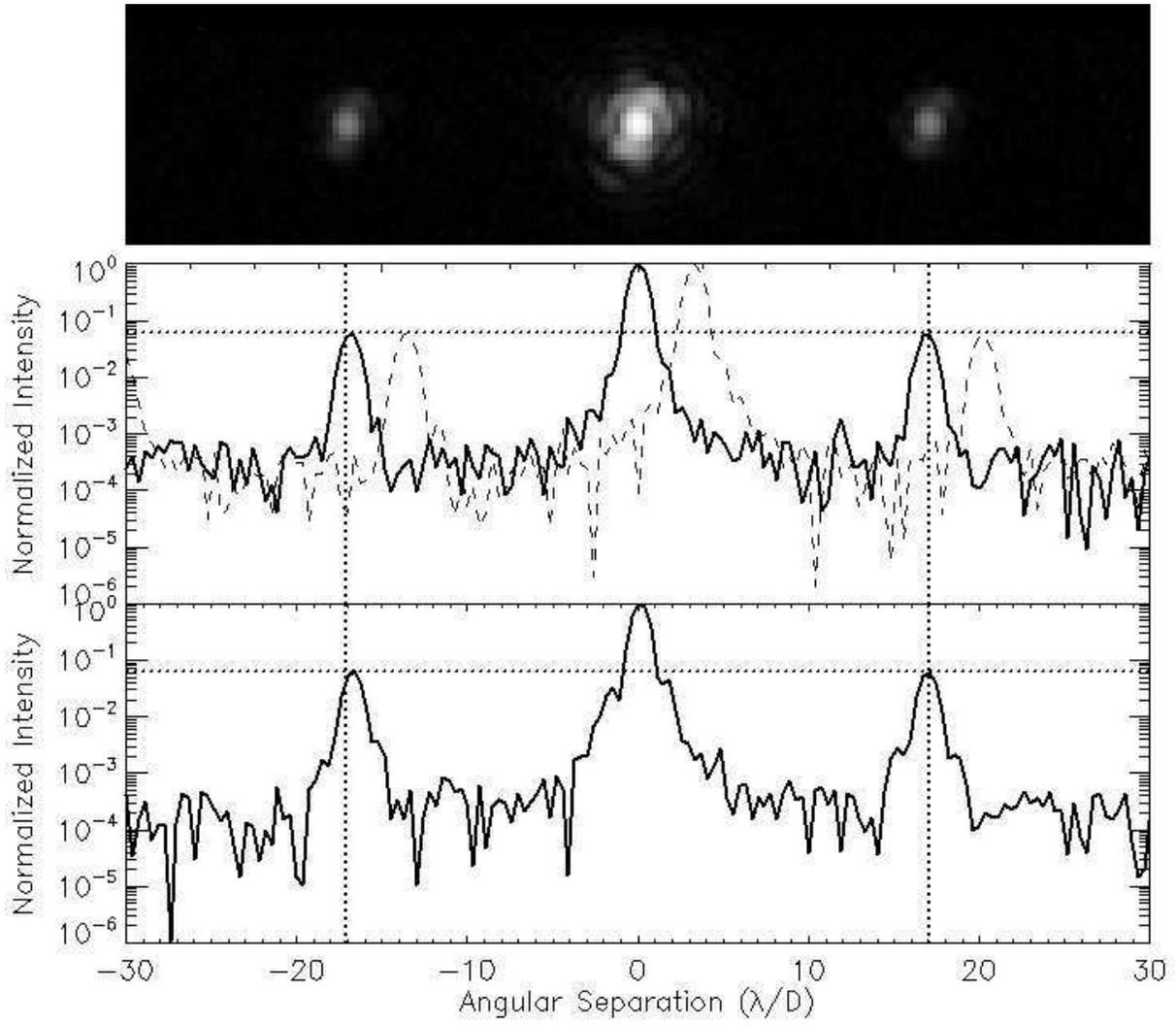}
\caption{Same as Fig.~\ref{fig1} for the laboratory experiment using a transmissive amplitude grating.\label{fig5}}
\end{figure}

A transmissive amplitude grating has been tested on the Universit\'{e} de Montr\'{e}al high contrast imaging testbed. Commercial lenses were used to make a f/31 imaging system, yielding, at 1.625~$\mu $m and with a 3\% filter bandpass, a PSF having 2.7~pixels per $\lambda/D$. The pupil diameter was 13.1~mm. A transmissive amplitude grating mask having $150\pm 10$~$\mu $m thick wires and $595\pm 15$~$\mu $m spacing was introduced 10~mm from the pupil plane. This slight offset was necessary due to the mask and pupil stop mounts. The geometric pupil diameter at the grating mask conjugated plane is calculated to be 12.75~mm. With this setup, the diffraction envelope of a single wire drops to $0.97$ of its peak at a separation of $17.1$~$\lambda/D$, the location of the first satellite PSF pair. Neglecting the satellite PSFs chromaticity, it is less than 0.6 $\lambda/D$, it is thus expected that this grating will produce satellite PSFs having a relative intensity of $0.97\times (150/595)^2 = 0.062 \pm 0.012$ at $17.1 \pm 0.5$~$\lambda/D$ separation. The pinhole was moved and an aberration phase plate was introduced in the beam, producing a Strehl ratio 25\% lower, to confirm that these satellite PSFs do track the PSF core position and intensity. All acquired images were unsaturated and reduced by subtracting a dark frame and dividing by a flat field. The two brightest satellite PSFs are 160 times brighter than the background noise. Fig.~\ref{fig5} shows the resulting PSFs. Satellite PSFs have a relative intensity of 0.058 and are situated at 16.9~$\lambda/D$ from the PSF center, both values being consistent to the calculated values to 1$\sigma$ accuracy, and they track the PSF core position and its intensity, in agreement with what was expected. In all above experiments, it was possible to register the PSF to less than 0.1 pixel precision (1/30 $\lambda/D$) with the average position of two symmetric satellite PSFs and the relative intensities are constant to several percent (see Table~\ref{tab3}), all consistent with the expected accuracy for satellite PSFs that are 100-1000 times the background noise (see Table~\ref{tab2}).

\section{On-Sky Implementation}
\begin{figure}
\epsscale{1}
\plotone{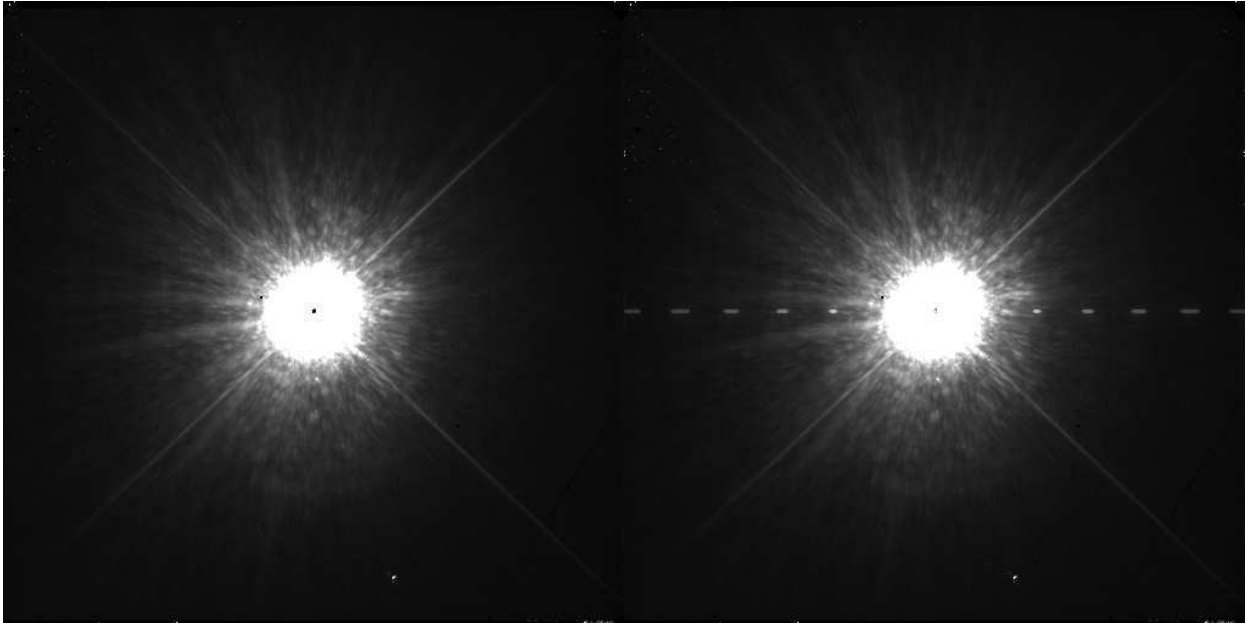}
\caption{Observed 30~s CH$_4$-band Gemini Altair/NIRI saturated PSF (left, star has $m_H$ = 5) and with simulated satellite PSFs (right). Simulated satellite PSFs are for a series of 42 wires having 18~$\micron $ thickness across a 84~mm pupil. Satellite PSFs are $\sim 10$ magnitudes fainter than the primary and are located every $\sim $1.7 arcsec interval along the X axis. They are produced using non-saturated data acquired in the same filter and normalized to the same integration time as the saturated image. Satellite PSFs have approximately the same intensity contrast in the FOV since the sinc modulation FWHM from the wire thickness (see Eq.~\ref{young}) is $\sim 9$ times the FOV (FOV is $22.4 \times 22.4$ arcsec). The satellite PSF elongation is simulated for the filter 6.5\% bandpass. PSF is saturated inside 0.8 arcsec radius.\label{fig6}}
\end{figure}

If we consider the Gemini Altair AO system \citep{herriot1998} and the NIRI near-infrared camera \citep{hodapp2000} for a possible implementation, satellite PSFs $\sim 10$ magnitudes fainter than the central peak should be $\sim $100 times brighter than the speckle noise at 4~arcsec \citep{marois2006}. If we put a transmission amplitude grating just in front of Altair's deformable mirror, where the wavefront diameter is 84~mm, a wire spacing and thickness of 2~mm and 18~$\mu$m would respectively be required to produce satellite PSFs 10 magnitudes fainter at 1.7 arcsec intervals. Such a configuration has 99.1\% transmission (see Fig.~\ref{fig6} for a simulated example).

For high contrast imaging seeking for detection limits of the order of 20 magnitudes at less than 1~arcsec, e.g the Gemini Planet Imager (GPI) and the VLT Planet Finder (VLTPF) projects, wire thickness will become an issue. A potential solution is to increase the wire spacing to place the satellite PSFs at 5~$\lambda/D$. Using the same 18~$\mu $m thick wires and wavefront diameter as the above example, it would produce satellite PSFs that are 15 magnitudes fainter than the primary. For more advanced coronagraphic projects aiming for Earth-like exoplanet detections at 24 magnitudes contrast, a simple approach is to put the wires directly on the primary mirror. A series of ten 18 microns thick wires across an 8-m diameter telescope will produce satellite PSFs 23.2 magnitudes fainter than the primary at 10 $\lambda/D$ (99.998\% transmission).

Another solution is to generate satellite PSFs using a small periodic phase aberration. In a conventional adaptive optics system with a separate wavefront sensor this could be achieved by adding a sinusoidal offset to the control point that the wavefront sensor attempts to achieve. Most AO systems operate with their wavefront control loop seeking not a flat wavefront in the sensor but a wavefront offset by the conjugate of any non-common-path aberrations between the wavefront sensor and science channel. Adding an additional sinusoidal component to this offset would result in a time-averaged PSF containing the satellite PSFs. As long as the sinusoid is lower in amplitude than the typical residual wavefront, 200-300 nm for current AO systems \citep{vandam2003} and 50-100~nm for extreme AO \citep{macintosh2004,mouillet2004}, and the wavefront sensor is operating in a linear regime this should have little effect on other components of the image. For systems without a separate wavefront sensor - e.g. the Terrestrial Planet Finder Coronagraph, which intends to use science-camera images on a reference target to control the deformable mirror via iterative phase diversity, the satellite images could be built into the target wavefront or focal-plane image that the phase diversity image sequence converges to. The satellite PSFs can then be used to estimate the PSF changes between the reference and science targets.

\section{Discussion}
Simulations presented in this paper are monochromatic. For bandpasses of finite size, off-axis satellite PSFs will be elongated and thus fainter relative to the core. The satellite PSF elongation is simply proportional to the filter bandpass $\Delta \lambda/\lambda$ times the satellite PSF angular separation. Additionally, for ground-based imaging, if no atmospheric dispersion corrector is used, the elongation of the satellite PSFs will not necessarily points towards the PSF center due to differential atmospheric refraction between the blue and red side of the bandpass \citep{filippenko1982,roe2002}. For H-band observations within $\pm 2$HA, the amplitude of this effect can be several tens of mas, i.e. of the order of one $\lambda/D$, as observed by precise registration of two near-infrared narrow-band PSFs made with the TRIDENT camera at CFHT \citep{marois2005}. However, satellite PSF chromaticity and differential atmospheric refraction do not influence the astrometry precision since the midpoint of the line connecting the centers of opposite satellite PSFs still coincides with the PSF center. For large bandpasses and/or satellite PSFs at large angular separations, the satellite PSF photometry can be partially correlated with a Strehl ratio variation since the required aperture necessary to estimate the satellite PSF intensity will need to be larger than a $\lambda/D$ diameter; it would thus be better to choose a satellite PSF at small separation to minimize this effect.

Distortions coming from optics after the added mask can change the relative position of the satellite PSFs. The satellite PSF position will thus need to be calibrated using an internal source or on-sky observations. To obtain precise relative astrometry of detected companions, distortion corrections for the entire optical system is also required. For ground-based imaging, two effects are potentially not correctable with the satellite PSF technique. If a wide bandpass is used, spectral differences between the companion and the primary coupled with differential atmospheric refraction can introduce small astrometric errors as a function of airmass. This effect is negligible for IFUs or narrowband imaging and can, for cases when it is important, be accounted for by analysis the companion spectrum. Anisoplanatism variations can also degrade the companion Strehl ratio without affecting that of the primary, thus possibly introducing companion photometric errors. Since high contrast imaging projects usually have small FOV (several arcsec wide) compared to standard AO anisoplanatism angle ($\sim $30~arcsec radius), such effect is relatively small.

The satellite PSFs will occupy a small portion of the FOV. Once the PSF is registered and flux normalized, they could be removed to avoid masking faint sources. Since the angular separation and size of the satellite PSFs is chromatic and fixed with respect to the instrument, a simple solution to remove them while conserving the flux of nearby sources would be to use the simultaneous spectral differential imaging SSDI \citep{racine1999,marois2000,sparks2002,biller2004,marois2005} and/or the angular differential imaging techniques ADI \citep{marois2004phd,liu2004,marois2006}. If such techniques still leave important localized residuals, the satellite PSFs can simply be masked.

\section{Conclusion}
It was shown both analytically and numerically that periodic amplitude and phase masks conjugated to the pupil plane with and without a coronagraph can be used to produce satellite PSFs that track the PSF core position and its intensity. The astrometric and photometric precision is proportional to the square root of the ratio of the satellite PSF intensity to the background speckle noise. Satellite PSFs 100 times brighter than the speckle noise achieve a registering and photometric accuracy of $0.3$ pixel RMS ($\sim $1/15 $\lambda/D$) and 5\% respectively. Such a precision is sufficient to properly combine images and to confirm proper motion of nearby systems within one year with AO on a 10-m telescope.

As illustrated by the laboratory experiment, this technique can be easily implemented by using a transmissive amplitude grating. The experiment has validated that an amplitude mask conjugated to the pupil plane produces such satellite PSFs and that they track the PSF core position and intensity while demonstrating the simplicity of the method.

This technique could play an important role in current AO systems and next generation extreme adaptive optics coronagraph systems like the Gemini Planet Imager \citep{macintosh2004,macintosh2006} or the VLT Planet Finder \citep{mouillet2004} as well as in future high contrast coronagraphic space observatories.

\acknowledgments
Authors would like to thank Anand Sivaramakrishnan for stimulating discussions. This research was performed under the auspices of the US Department of Energy by the University of California, Lawrence Livermore National Laboratory under contract W-7405-ENG-48, and also supported in part by the National Science Foundation Science and Technology Center for Adaptive Optics, managed by the University of California at Santa Cruz under cooperative agreement AST 98-76783. This work is supported in part through grants from the Natural Sciences and Engineering Research Council, Canada and from the Fonds Qu\'{e}b\'{e}cois de la Recherche sur la Nature et les Technologies, Qu\'{e}bec.

%\clearpage

\begin{center}
\begin{table}
\begin{center}
\caption{Astrometric and Photometric Precision with Mask Offset ($\lambda/D$ = 4 pixels, no aberration).\label{tab0}}
\begin{tabular}{cccccc}\hline
Mask type & Offset (period) & Coro. & $\Delta $xc & $\Delta $yc & $I_{\rm{satel}}/I_{\rm{PSF}}$\\ \hline
Phase     &     0          &  no   &     -0.31        &    0.00         &      0.00100        \\
          &     $1/4$    &  no   &      0.00       &     0.00        &       0.00099            \\
          &     0          &  yes  &      0.00     &      0.00       &     0.00100            \\
          &     $1/4$    &  yes  &      0.00       &     0.00        &       0.00099             \\
Amplitude &     0          &  no   &      0.00       &    0.00         &       0.00100         \\
          &     $1/4$    &  no   &      0.00   &     0.00        &          0.00088                 \\
          &     0          &  yes  &      0.00       &    0.00         &       0.00100               \\
          &     $1/4$    &  yes  &      0.00       &    0.00         &     0.00100                  \\
Grating   &     0          &  no   &      0.00   &    0.00        &        0.01211                   \\
          &     $1/4$    &  no   &      0.00       &   0.00          &    0.01237              \\
          &     0          &  yes  &      0.00       &   0.00          &    0.01240                \\
          &     $1/4$    &  yes  &      0.00       &   0.00          &     0.01240                  \\
\hline
\end{tabular}
\end{center}
\end{table}
\end{center}
\clearpage

\clearpage
\begin{center}
\begin{table}
\begin{center}
\caption{Simulated Astrometric and Photometric Precision ($\lambda/D$ = 4 pixels).\label{tab1}}
\begin{tabular}{lccccccc}\hline
Aberration  & Coro. &    With   & $\Delta $XC rms & $\Delta $YC rms & $I_{\rm{satel}}/I_{\rm{PSF}}$ & $I_{\rm{satel}}/I_{\rm{PSF}}$ & $I_{\rm{satel}}/I_{\rm{PSF}}$\\ 
        &        & 150nm RMS&  (pixel)  &  (pixel)  &                      &  rms (\%) & expected\\ \hline
Phase    & no & no  &  - &   -    & 0.00100&     -   & 0.00100\\
Phase    & no & yes &   0.19&   0.18    & 0.00100& 0.00001 (1.0)& 0.00100\\
Phase    & yes& yes &   0.15&   0.10    & 0.00099& 0.00001 (1.0)& 0.00100\\
Amplitude& no & no  &   - &   -    & 0.00100&     -   & 0.00100\\
Amplitude& no & yes &   0.24&   0.20    & 0.00102& 0.00005 (5)& 0.00100\\
Amplitude& yes& yes &   0.19&   0.18    & 0.00103& 0.00005 (5)& 0.00100\\
Grid     & no & no  &   - &   -    & 0.01240&     -   & 0.01235\\
Grid     & no & yes &   0.12&   0.15    & 0.01198& 0.00010 (0.8)& 0.01235\\
Grid     & yes& yes &   0.05&   0.03    & 0.01235& 0.00010 (0.8)& 0.01235\\
%Phase    & no & no  &    -  &    -      & 0.00027&     -   & 0.00020\\
%Phase    & no & yes &   0.06&   0.05    & 0.00021& $9.1\times 10^{-6}$ (4.3)& 0.00020\\
%Phase    & yes& yes &   0.01&   0.01    & 0.00020& $6.7\times 10^{-6}$ (3.4)& 0.00020\\
%Amplitude& no & no  &    -  &    -      & 0.00020&     -   & 0.00020\\
%Amplitude& no & yes &   0.11&   0.07    & 0.00020& $1.3\times 10^{-6}$ (0.6)& 0.00020\\
%Amplitude& yes& yes &   0.07&   0.06    & 0.00020& $1.2\times 10^{-6}$ (0.6)& 0.00020\\
%Grid     & no & no  &    -  &    -      & 0.01197&     -   & 0.01235\\
%Grid     & no & yes &   0.02&   0.01    & 0.01197& $5.8\times 10^{-6}$ (0.5)& 0.01235\\
%Grid     & yes& yes &   0.01&   0.01    & 0.01234& $4.4\times 10^{-6}$ (0.4)& 0.01235\\
\hline
\end{tabular}
\end{center}
\end{table}
\end{center}
\clearpage

\clearpage
\begin{center}
\begin{table}
\begin{center}
\caption{Satellite PSF Intensity vs Astrometric and Photometric Precision ($\lambda/D$ = 4 pixels).\label{tab2}}
\begin{tabular}{cccccc}\hline
Satellite PSF int.& Coro. & $\Delta $XC rms & $\Delta $YC rms & $I_{\rm{satel}}/I_{\rm{PSF}}$ & $I_{\rm{satel}}/I_{\rm{PSF}}$ \\
 to speckle noise ratio$^{\rm{a}}$            &       &    &    &                      &               rms (\%)      \\ \hline
  10     & no & 0.78& 0.93& 0.00013& 0.00002 (15)\\
  16     & yes& 0.75& 0.80& 0.00011& 0.00002 (18)\\
  100    & no & 0.24& 0.20& 0.00102& 0.00005 (5)\\
  160    & yes& 0.19& 0.18& 0.00103& 0.00005 (5)\\
  1,000  & no & 0.07& 0.06& 0.01003& 0.00018 (1.8)\\
  1,600  & yes& 0.06& 0.05& 0.01005& 0.00014 (1.4)\\
  10,000 & no & 0.02& 0.02& 0.10214& 0.00092 (0.9)\\
  16,000 & yes& 0.02& 0.02& 0.10156& 0.00067 (0.7)\\
% & 10     & no & 0.98& 0.38& 0.00003& $8.6\times 10^{-7}$ (3.4)\\
% & 10     & yes& 0.22& 0.19& 0.00002& $4.6\times 10^{-7}$ (2.3)\\
% & 100    & no & 0.11& 0.07& 0.00020& $1.2\times 10^{-6}$ (0.6)\\
%30& 100    & yes& 0.07& 0.06& 0.00020& $1.2\times 10^{-6}$ (0.6)\\
% & 1,000  & no & 0.03& 0.02& 0.00201& $3.7\times 10^{-6}$ (0.2)\\
%& 1,000  & yes& 0.02& 0.02& 0.00200& $3.4\times 10^{-6}$ (0.2)\\
%& 10,000 & no & 0.01& 0.01& 0.02029& $1.9\times 10^{-5}$ (0.1)\\
% & 10,000 & yes& 0.01& 0.01& 0.02008& $1.4\times 10^{-5}$ (0.1)\\
\hline
\end{tabular}
\end{center}
{\footnotesize $^{\rm{a}}$ In all cases, satellite PSF intensities are 0.1\% fainter than the PSF peak intensity. The satellite PSF intensity to speckle noise ratios are higher when using a coronagraph since, for these cases, the speckle noise is attenuated by the coronagraph by a factor 1.6 in our simulations.}
\end{table}
\end{center}
\clearpage

\begin{center}
\begin{table}
\begin{center}
\caption{Laboratory Experiment.\label{tab3}}
\begin{tabular}{cccccccc}\hline
 & \multicolumn{2}{c}{PSF} & \multicolumn{2}{c}{Satellite PSFs} & & & \\
 & XC & YC & XC & YC & $\Delta $XC & $\Delta $YC & $<I_{\rm{satel}}/I_{\rm{PSF}}>$ \\ \hline
Unaberrated & 199.93 & 199.98 & 199.90 & 200.05 & 0.03 & -0.07 & 0.055 \\
Aberrated & 200.08 & 199.69 & 199.99 & 199.68 & 0.09 & 0.01 & 0.061 \\
Shifted & 200.45 & 209.04 & 200.45 & 209.07 & 0.00 & -0.03 & 0.059 \\ \hline
\end{tabular}
\end{center}
\end{table}
\end{center}
\clearpage

\clearpage

\clearpage

\clearpage

\clearpage

\clearpage

\clearpage

\end{document}